\documentclass[showpacs,aps,superscriptaddress, twocolumn,amsmath]{revtex4}
\usepackage{txfonts}
\usepackage{color}
\usepackage{amssymb}
\usepackage{graphicx}
\usepackage{dcolumn}
\usepackage{bm}
\usepackage{hyperref}
\begin{document}
\preprint{APS/123-QED}
\title{Novel Two-dimensional SiC$_2$ Sheet with Full Pentagon Network}
\author{J.  Liu}
\affiliation{Department of Physics, Xiangtan University, Xiangtan 411105, China}
\author{C. Y.  He}
\affiliation{Department of Physics, Xiangtan University, Xiangtan 411105, China}
\author{N. Jiao}
\affiliation{Department of Physics, Xiangtan University, Xiangtan 411105, China}
\author{H. P. Xiao}
\email{hpxiao@xtu.edu.cn}
\affiliation{Department of Physics, Xiangtan University, Xiangtan 411105, China}
\author{K. W. Zhang}
\affiliation{Department of Physics, Xiangtan University, Xiangtan 411105, China}
\author{R. Z. Wang}
\affiliation{College of Materials Science and Engineering, Beijing University of Technology, Beijing 100124, China}
\author{L. Z. Sun}
\email{lzsun@xtu.edu.cn}
\affiliation{Department of Physics, Xiangtan University, Xiangtan 411105, China}

\date{\today}
\begin{abstract}
We propose a promising two-dimensional nano-sheet of SiC$_2$ (SiC$_2$-pentagon) consisting of tetrahedral silicon atoms and triple-linked carbon atoms in a fully-pentagon network. The SiC$_2$-pentagon with buckled configuration is more favorable than its planar counterpart and previously proposed SiC$_2$-silagraphene with tetra-coordinate silicon atoms; and its dynamical stability is confirmed through phonon analyzing. Buckled SiC$_2$-pentagon is an indirect-band-gap semiconductor with a gap of 1.388 eV.  However, its one-dimensional nanoribbons can be metals or semiconductors depending on the edge type, shape, and decoration. Finally, we propose a method to produce the buckled SiC$_2$$-$pentagon through chemical exfoliation on the $\beta$$-$SiC(001)-c$(2\times2)$ SDB surface.\\
\end{abstract}
\maketitle
\section{Introduction}
\indent Group 4 elements have versatile stable and metastable phases due to their ability to form sp-, sp$^2$- and sp$^3$-hybridized chemical bonds. Graphene\cite{1}, a single layer of sp$^2$-hybridized carbon atoms, is considered as a promising candidate for future nanoelectronics\cite{2,3} in views of its unique electronic properties such as super high carrier mobility \cite{4,5} and anomalous quantum hall effect \cite{6,7} derived from its linear energy dispersion \cite{8,9,10}. Silicon is the most fundamental material in traditional electronic industry. Its two-dimensional (2D) allotrope silicene have been theoretically investigated\cite{11, 12, 13, 14, 15} and fabricated in experiments recently \cite{16, 17, 18, 19, 20, 21}. Silicene possessing the same crystal lattice as graphene is considered as a powerful competitor for graphene in future nano-electronics due to its many excellent electronic properties. Especially, silicene system possesses relatively stronger spin-orbit coupling (SOC) \cite{15} in comparison with graphene providing us a possible quantum-spin-hall-effect system. Unfortunately, silicene dose not the ground state of silicon. Under ambient condition, sp$^2$-hybridized graphene is more favorable than the three-dimensional (3D) sp$^3$-hybridized diamond and acknowledged as the ground state of carbon but silicon prefers its 3D cubic phase (diamond-structure) than the 2D silicene. Usually, the experimentally produced silicenes prefer sp$^2$-sp$^3$ mixed configuration than the graphene-liked honeycomb lattice. The experimental synthesis of perfect silicene is still a challenge for current technology.\\
\indent The theoretical predictions of new materials are very important for their discovery. Many novel materials are theoretically predicted before their experimental fabrication, such as silicene\cite{11, 12, 13, 14, 16, 17, 18} and graphane \cite{22, 23, 24}. Theoretically, we can evaluate the probability and the possibility of the existence of a new material through simulating its thermodynamical stability and dynamical stability, respectively. It is also significant to propose and design the potential approaches to synthesize it in experiments. In addition to the pure carbon graphene allotropes and pure silicene, researchers recently proposed many 2D binary compounds of carbon and silicon such as the SiC sheet \cite{25, 26, 27, 28, 29} with the same lattice as boron nitride sheet and the SiC$_2$-silagraphene\cite{30} with tetra-coordinate silicon atoms.\\
\indent In present work, using the segment combination method, we propose a promising two-dimensional (2D) nano-sheet of SiC$_2$ (buckled SiC$_2$-pentagon) through combination of tetrahedral silicon atoms and triple-linked carbon atoms in a fully-pentagon networks locating at the four-linked nodes and the three-linked nodes, respectively. The dynamical stability and the electronic structure of the buckled SiC$_2$-pentagon are evaluated by ab initio calculations. We find that the buckled SiC$_2$-pentagon is more favorable one than the previously proposed planar-SiC$_2$-silagraphene and its dynamical stability is confirmed. The electronic properties of the one-dimensional nanoribbons based on the buckled SiC$_2$-pentagon are systematically investigated. Interestingly, we find that SiC$_2$-pentagon nanosheet has the similar structure with the top layers of the $\beta$-SiC(001)-$_{c}$($2\times2$) SDB surface \cite{31}.  Referring to the successful preparation of graphene by mechanical exfoliation, we propose that SiC$_2$-pentagon nanosheet can be obtained through chemical exfoliation method from the $\beta$-SiC(001)-$_ c$$(2\times2)$ SDB surface.\\
\section{Models and Computational Methods}
\begin{figure}
  \includegraphics[width=3.2in]{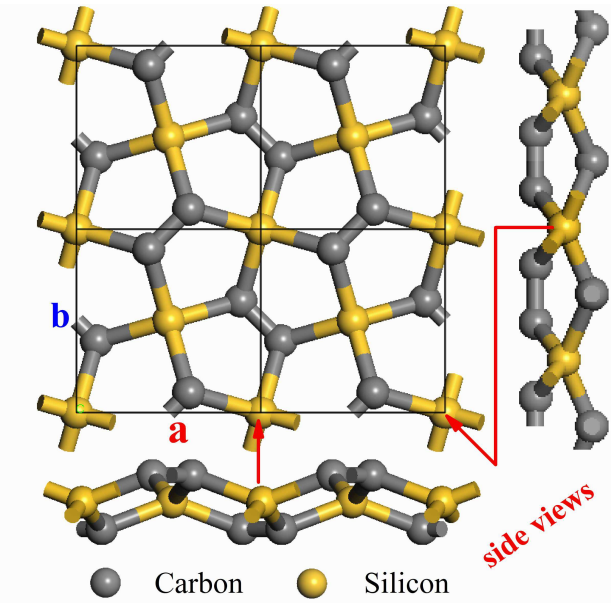}
  \caption{Schematic crystal structure of the buckled SiC$_2$-pentagon.} \label{fig1}
\end{figure}

\indent The hexagonal honeycomb-like lattice of graphene and silicene is a fully-hexagon network where every node is equivalent and triple-linked. Based on fully pentagonal network, which contains both four- and triple-linked nodes, we propose a fully-pentagon rule-breaking 2D nanosheet through the combination of the ground states of sp3-hybridized Silicon and sp2-hybridized Carbon locating at the four- and triple-linked nodes, respectively, as shown in \ref{fig1}. Such segment combination method proposed in our previous work\cite{segment} has successfully predicted many carbon allotropes based on the most and second stable cubic-diamond and hexagonal-diamond. Both planar and buckled conditions are considered, and the composition ratio of silicon and carbon atoms is 1: 2 .  We name them planar-SiC$_2$-pentagon and buckled-SiC$_2$-pentagon, respectively. The previously proposed planar-SiC$_2$-silagraphene\cite{30} and its buckled version are also considered in our present work for comparison purpose in views of the fact that they possess the same topological units of four- and triple-linked nodes with the same ration of 1:2. However, SiC$_2$-silagraphenes are topologically different from SiC$_2$-pentagon, SiC$_2$-silagraphene contain both hexagons and rhombuses and SiC$_2$-pentagon contain only pentagons.\\
\indent To evaluate the stability and electronic structures of SiC$_2$-pentagon, we adopt the Vienna Ab initio Simulation Package (VASP)\cite{vasp} based on density functional theory (DFT) to perform the first-principles plane-wave calculations for its structural optimizations, total energy calculations, and electronic structures. The projected augmented wave (PAW)\cite{paw} potentials are chosen to describe the electron-ionic core interaction and the PBE version of the generalized gradient approximation (GGA)\cite{gga} is adopted for the electronic exchange and correlation. A plane-wave basis set with the kinetic energy cutoff of 500 eV is employed. The Brillouin Zone sample meshes based on the Monkhorst-Pack scheme are set to be denser enough (with k-points separations less than 2$\pi$$\times$0.04 {\AA}$^{-1}$) for all systems to ensure our calculation precision. Lattice constants and atom positions for all systems considered in present work are fully optimized until the residual force on each atom to be less than 0.001 eV/{\AA}. To evaluate the dynamical stability of the 2D SiC$_2$ compounds, we calculate their phonon band structures and phonon density of states (DOS) using the phonon\cite{phonon} package with applying forces from VASP calculations. The energetic stability is evaluated through comparing the cohesive energies.\\
\section{Results and discussions}
\begin{figure}
 \includegraphics[width=3.3in]{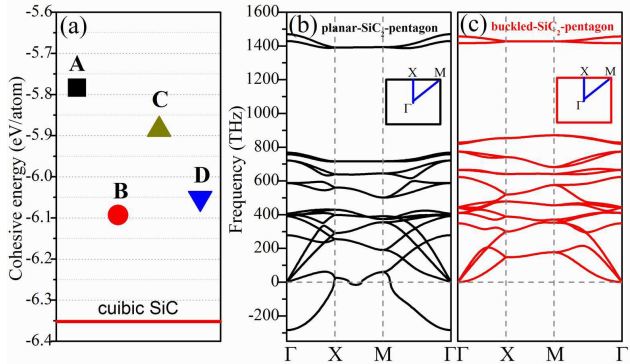}\\
  \caption{(a) Cohesive energy for the four 2D SiC$ _2$ configrations, the cohesive of cubic SiC is also included for comparison purpose.  A, B, C, and D in (a) denote planar-SiC$_2$-pantagon, buckled-SiC$_2$-pentagon, planar-SiC$_2$-silagraphene, and buckled-SiC$_2$-silagraphene, respectively.  (b) and (c) are the phonon bands structures for planar-SiC$_2$-pentagon and buckled-SiC$_2$-pentagon, respectively.}
  \label{fig2}
\end{figure}

\indent We first introduce the fundamental structural information about these 2D SiC$_2$ sheets. Planar-SiC$_2$-pentagon possesses a tetragonal lattice with constants of
a=b=4.735 {\AA}, c=20{\AA}. It belongs to P4/mbm (127) space group and possesses two inequivalent atomic positions of (0.0, 0.0, 0.5) and (0.6, 0.1, 0.5) for silicon and carbon atoms, respectively. Silicon and carbon atoms are located in the same plane. There are two inequivalent chemical bonds in planar SiC2-pentagon, namely the C-C and Si-C bonds, with length of 1.344 {\AA} and 1.951 {\AA}, respectively. And three inequivalent bond angles are $\angle$Si-C-Si=118.18$^\circ$, $\angle$Si-C-C=120.91$^\circ$, and $\angle$C-Si-C=90$^\circ$, respectively. The buckled-SiC$_2$-pentagon possesses a tetragonal lattice with constants of a=b=4.408 {\AA}, c=20{\AA}. It belongs to P-421m (113) space group and possesses two inequivalent atomic positions of (0.0, 0.0, 0.5) and (0.609, 0.109, 0.533) for silicon and carbon atoms, respectively. Its two inequivalent chemical bonds are the C-C and Si-C bonds with the length of 1.362 {\AA} and 1.908 {\AA}, respectively. Four inequivalent bond angles are $\angle$Si-C-Si=109.621$^\circ$, $\angle$Si-C-C=117.409$^\circ$, intra-sheet $\angle$C-Si-C=96.9$^\circ$ and outer-sheet $\angle$C-Si-C=139.43$^\circ$, respectively. The layer thickness of  buckled-SiC$_2$-pentagon is 1.32 {\AA}.\\
\indent Planar-SiC$_2$-silagraphene possesses an orthorhombic lattice with constants of a=2.809 {\AA}, b=3.925 {\AA} and c=20 {\AA}. It belongs to Pmmm (47) space group and possesses two inequivalent atomic positions of (0.0, 0.0, 0.5) and (0.5, 0.331, 0.5) for silicon and carbon atoms, respectively. The two inequivalent chemical bonds, namly the C-C and Si-C bonds, in planar-SiC$_2$-silagraphene are 1.329 {\AA} and 1.912 {\AA}, respectively. Four inequivalent bond angles are $\angle$Si-C-Si=94.544$^\circ$, $\angle$Si-C-C=132.73$^\circ$, intra-hexagon $\angle$C-Si-C=94.544$^\circ$ and intra-rhombus $\angle$C-Si-C=85.456$^\circ$, respectively. Buckled-SiC$_2$-silagraphene possesses an orthorhombic lattice with constants of a=6.219 {\AA}, b=8.049 {\AA}, and c=20 {\AA}. It belongs to Cmma (67) space group and possesses two inequivalent atomic positions of (0.25, 0.0, 0.5) and (0.5, 0.158, 0.541) for silicon and carbon atoms, respectively. The length of the inequivalent C-C and Si-C bonds in buckled-SiC$_2$-silagraphene are  1.487 {\AA} and 2.170 {\AA}, respectively. Five inequivalent bond angles are $\angle$Si-C-Si=91.527$^\circ$, $\angle$Si-C-C=125.786$^\circ$, intra-hexagon $\angle$C-Si-C=108.429$^\circ$, intra-rhombus $\angle$C-Si-C=88.473$^\circ$, and outer-sheet $\angle$C-Si-C=135.277$^\circ$, respectively. The layer thickness of buckled-SiC$_2$-silagraphene is 1.64 {\AA}. From the analysis of the structural information of those $SiC_{2}$, we can see that the length of the C-C bond in the buckled $SiC_{2}$ is larger than that in the planar ones. Moreover, from the point of view of the configuration and bond length of $SiC_{2}$, we can see that carbon atom dose not form a standard sp$^2$-hybridization and silicon atom dose not form a standard sp$^3$-hybridization. \\
\indent To evaluate the relative stability of these four configurations, we calculate their cohesive energies for each atom as:
\begin{eqnarray}\label{equ1}
E_{coh}=\frac{E_{tot}-n\times E_{C}-m\times E_{Si}}{n+m}
\end{eqnarray}
Where $E_{tot}$, $E_{C}$, and $E_{Si}$ represent the total energies of the $SiC_{2}$, isolated C atom, and isolated Si atom, respectively. $n$ and $m$ are the numbers of C and Si atoms, respectively. As shown in \ref{fig2} (a), we can see that the cohesive energies of the planar SiC$_2$-pantagon, buckled SiC$_2$-pentagon, planar SiC$_2$-silagraphene and buckled SiC$_2$-silagraphene are -5.783 eV/atom, -6.093 eV/atom, -5.885 eV/atom and -6.052 eV/atom, respectively. Although the cohesive energy of the four 2D SiC$_2$ sheets is larger than that of cubic SiC (-6.352), the stability of both buckled SiC$_2$-pentagon and buckled SiC$_2$-silagraphene are more closer to the reference. From the point of view of cohesive energy,  buckled SiC$_2$-pentagon is the most stable one in the four 2D SiC$_2$ sheets . The cohesive energy of buckled SiC$_{2}$-pentagon structure is 0.04eV per atom smaller than that of buckled $SiC_{2}$-silagraphene. Moreover, the results also indicate that silicon prefer the tetrahedral configuration than the tetracoordinate one according to the fact that the buckled structures are more favorable than planar ones.\\
\indent To investigate the dynamic stability of planar-SiC$_2$-pentagon and buckled-SiC$_2$-pentagon systems, we calculate their phonon band structures, as shown in
\ref{fig2} (b) and (c). The phonon spectrum is obtained along the high symmetry direction of the Brillouin zone of the systems with the point of G, X, M. From this picture, we can see that buckled-SiC$_2$-pentagon obviously has no imaginary frequency in the phonon band structures, as shown in \ref{fig2} (c). Moreover, the phonon density of states (DOS) further confirms such conclusion. The results indicate that the buckled-SiC$_2$-pentagon is dynamically stable. However, the phonon band structures of planar-SiC$_2$-pentagon show imaginary frequency which indicates it is dynamically unstable. We also calculate the phonon band structures of planar-SiC$_2$-silagraphene and buckled-SiC$_2$-silagraphene and the results also show that only buckled-SiC$_2$-silagraphene is dynamically stable. The results indicate that buckled configuration of 2D-SiC$_2$ is more favorable for such systems. The above discussions show that the most promising one among these 2D SiC$_2$ sheets is the buckled-SiC$_2$-pentagon. Then we consider if there is a potential approach to synthesis such a novel nano-material. From the point of view of structure, we find that the lattice constants (a=b=4.408 {\AA}) of buckled-SiC$_2$-pentagon are very close to that of the (100) surface of cubic SiC. Especially, the buckled-SiC$_2$-pentagon has the similar structure with the $\beta$$-$SiC(001)-c$(2\times2)$ SDB surface\cite{31}. If the outermost layers of $\beta$$-$SiC(001)-c$(2\times2)$ SDB surface can be striped out, it will form buckled-SiC2-pentagon. Such issue will be discussed in detail below.\\
\begin{figure}
  \includegraphics[width=3.3in]{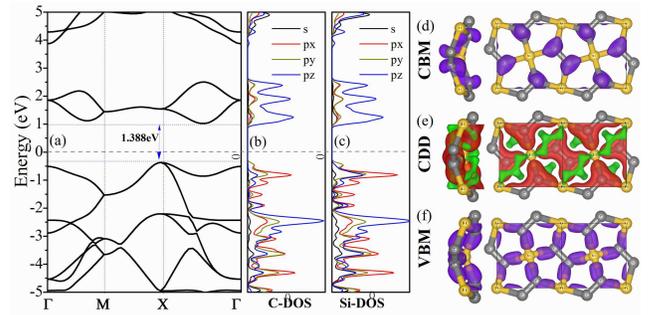}\\
  \caption{(a), (b), and (c) are band structure, PDOS-C, and PDOS-Si for buckled-SiC$_2$-pentagon, respectively. (d) and (f) are the charge densities of VBM and CBM of buckled-SiC$_2$-pentagon, respectively. (e) is the charge density difference (CDD) for buckled-SiC$_2$-pentagon, where red and green isosurfaces indicate charge accumulation and depletion regions, respectively.}
  \label{fig3}
\end{figure}
\indent We then study the electronic properties of the buckled-SiC$_2$-pentagon and its one-dimensional (1D) ramifications. The band structure of the buckled-SiC$_2$-pentagon is shown in \ref{fig3} (a) and the density of states projected on carbon-atomic-orbits (PDOS-C) and silicon-atomic-orbits (PDOS-Si) are shown in \ref{fig3} (b) and (c), respectively. The band structures show that buckled-SiC$_2$-pentagon is an indirect-band gap semiconductor with a gap of 1.388 eV. From the PDOSs, we can see that carbon atom dose not form a standard sp$^2$-hybridization and silicon atom dose not form a standard sp$^3$-hybridization due to the mixed existence of the states of px, py, and pz around the valence band maximum (VBM) and conduction band minimum (CBM), which agree well with the above analysis. The formants in both PDOS-C and PDOS-Si show strong interactions between carbon and silicon atoms and the formation of chemical Si-C bonds. Such bonding effect can also be observed from the charge density difference (CDD, defined as the difference between the total charge density in the system and the superpositions of neutral atomic charge densities placed at atomic sites) as shown in \ref{fig3} (e). The results indicate that the electrons transfer from silicon atoms in tetrahedral sp$^3$-like configuration to carbon atoms in a sp$^2$-like configuration. The corresponding charge densities of the VBM and the CBM are shown in \ref{fig3} (d) and (f), respectively. We can see the most active valence electrons (VBM electrons) in buckled-SiC$_2$-pentagon mainly distribute on the Si-C chemical bonds. The CBM states mainly distribute on carbon atoms along the direction perpendicular to the sheet indicating that it is mainly contributed by the anti-bonding pz state of carbon. Based on the analysis of PDOSs and the decomposed charge densities we can illustrate the bonding characteristics of buckled-SiC$_2$-pentagon as follows: buckled-SiC$_2$-pentagon contains only $\sigma$-type Si-C bonds and C-C bonds but no $\pi$-type C-C bonds formed by the p$_z$ state electrons of carbon. The p$_z$ state electrons of carbon are mixed into $\sigma$-type Si-C bonds. Each $\sigma$-type Si-C bond contains mixed 2s, 2p$_x$, 2p$_y$ and 2p$_z$ states of silicon as well as 2s, 2p$_x$, 2p$_y$, 2p$_z$ state electrons of carbon. Each $\sigma$-type C-C bond contains mixed 2s, 2p$_x$ and 2p$_y$ electrons of carbon. In view of missing the delocalized $\pi$-type electrons, buckled-SiC$_2$-pentagon behaves as semiconductor.\\
\begin{figure}
\includegraphics[width=3.5in]{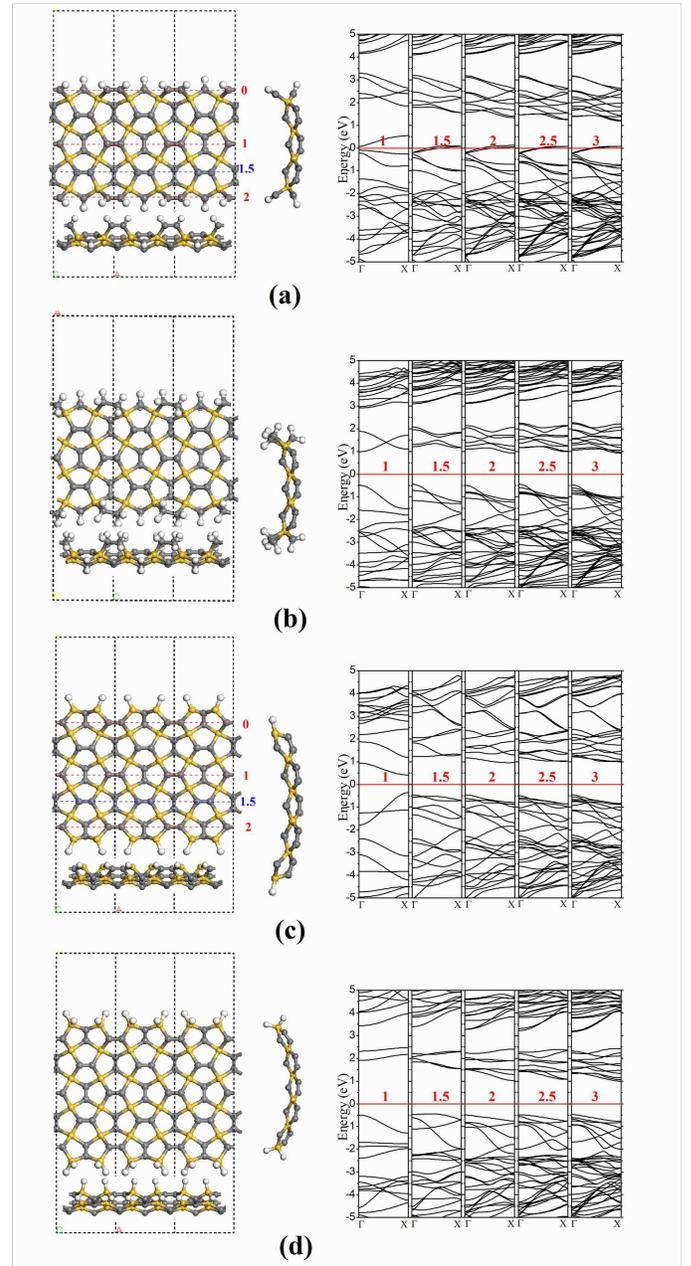}\\
 \caption{Configurations and band structures for nanoribbons of buckled-SiC$_2$-pentagon-CH$_2$ (a), buckled-SiC$_2$-pentagon-CH (b), buckled-SiC$_2$-pentagon-SiH (c) and buckled-SiC$_2$-pentagon-SiH$_2$ (d), respectively. }
 \label{fig4}
\end{figure}
\begin{figure}
\includegraphics[width=3.5in]{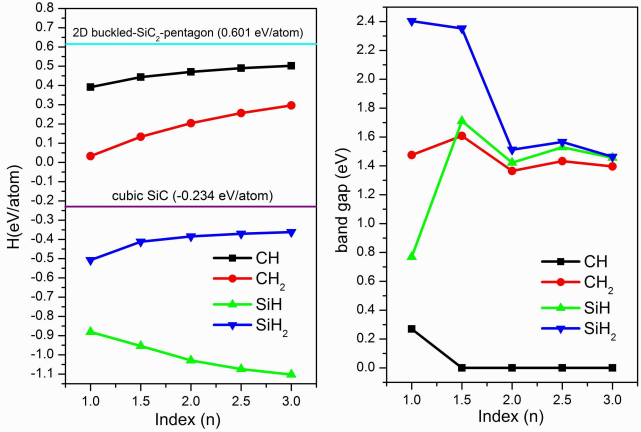}\\
 \caption{ Gibbs free energy (H) (a) and bad gaps (b) of buckled-SiC$_2$-pentagon nanoribbons with different edges and hydrogenations. CH and CH$_2$ denote carbon terminated ribbons with mono- and the di-hydrogenated edges, respectively. SiH and SiH$_2$ denote silicon terminated ribbons with mono- and the di-hydrogenated edges, respectively.}
 \label{fig5}
\end{figure}
\indent Based on the 2D buckled-SiC$_2$-pentagon sheet, we construct four groups of nanoribbons with different edge shapes and hydrogenations, named buckled-SiC$_2$-pentagon-CH, buckled-SiC$_2$-pentagon-CH$_2$, buckled-SiC$_2$-pentagon-SiH, and buckled-SiC$_2$-pentagon-SiH$_2$, as shown in \ref{fig4} (a), (b), (c) and (d), respectively. Each group contains five nanoribbons with different width (1, 1.5, 2, 2.5 and 3, the definition of the ribbon width is denoted in the figures). \ref{fig5} (a) shows the Gibbs free energies as functions of ribbon widths for the four groups of nanoribbons. The Gibbs free energy of formation $\delta$$G$ is defined as:\cite{gibbs}
\begin{equation}\label{equ1}
    \delta{G}=E(tot)-\chi_H\mu_H-\chi_C\mu_C-\chi_{Si}\mu_{Si}
\end{equation}
where $E(tot)$ is the cohesive energy per atom of the nanoribbon, $\chi_i$ is the molar fraction of atom $i$ ($i=H, C, Si$) in the structure satisfying $\chi_H+\chi_C+\chi_{Si}=1$, and $\mu_i$ is the chemical potential of each constituent atoms. We choose $\mu_H$ as the binding energy per atom of the $H_2$ molecular, $\mu_C$ and $\mu_{Si}$ as the cohesive energies per atom of a single layer graphene sheet and cubic silicon, respectively. The results of Gibbs free energy indicate that the 1D derivatives are always more favorable than the 2D buckled-SiC$_2$-pentagon because the 1D derivatives provide enough space to release the inner-sheet strains and reduce the energy of the system. Different ribbons will release different energies depending on their edge types, decorations as well as ribbons widths. We can see that buckled-SiC$_2$-pentagon-SiH and buckled-SiC$_2$-pentagon-SiH$_2$ with silicon termination are always more favorable than buckled-SiC$_2$-pentagon-CH and buckled-SiC$_2$-pentagon-CH$_2$ with carbon termination. Buckled-SiC$_2$-pentagon-SiH and buckled-SiC$_2$-pentagon-SiH$_2$ are even more favorable than the 3D cubic SiC. The Gibbs free energy of buckled-SiC$_2$-pentagon-SiH decrease as the increase in ribbon width, whereas the Gibbs free energy of buckled-SiC$_2$-pentagon-SiH$_2$ increase with the increasing of the ribbon width. As for carbon terminated systems, Gibbs free energy for both buckled-SiC$_2$-pentagon-CH and buckled-SiC$_2$-pentagon-CH$_2$ increases with the increasing of the ribbon width. Moreover, the Gibbs free energy for both buckled-SiC$_2$-pentagon-CH and buckled-SiC$_2$-pentagon-CH$_2$  is higher than that of 3D cubic SiC. The band structures of buckled-SiC$_2$-pentagon-CH, buckled-SiC$_2$-pentagon-CH$_2$, buckled-SiC$_2$-pentagon-SiH, and buckled-SiC$_2$-pentagon-SiH$_2$ are shown in \ref{fig4}. We can see that all of them are semiconductors except for the group of buckled-SiC$_2$-pentagon-CH. Buckled-SiC$_2$-pentagon-CH have similar edge shape and decoration like that of zigzag graphene nanoribbons (ZGNR), they possess similar metallic edge states with those of ZGNRs except for the ultra-narrow one with width of 1, which is a semiconductor with band gap of 0.27 eV. The band gaps of all the 1D ribbons are summarized in \ref{fig5} (b). We can see that the band gaps of the three semiconducting groups of ribbons distribute in the energy range of 0.76-2.43 eV and approach to 1.46 eV as the increase in the ribbon width. The results of the buckled-SiC$_2$-pentagon nanoribbons indicate that the electronic structures of buckled-SiC$_2$-pentagon can be effectively modulated by patterning them to quasi-one-dimensional materials.\\
\begin{figure}
\includegraphics[width=3.5in]{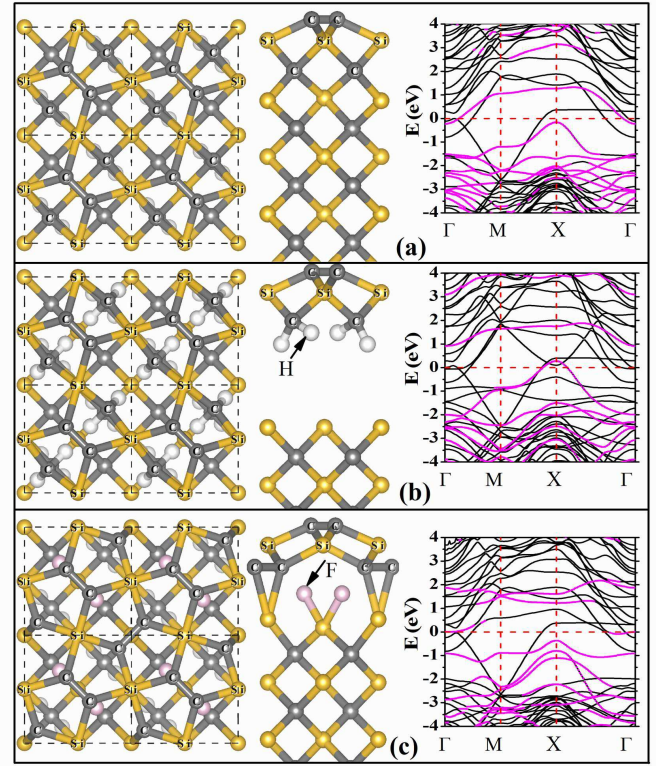}\\
 \caption{Equilibrium structure and band structures for (a) $\beta$$-$SiC(001)-c$(2\times2)$ SDB reconstruction, $\beta$$-$SiC(001)-c$(2\times2)$ SDB surface with (b) H and (c) F implantation. }
 \label{fig6}
\end{figure}
\indent Finally, from the point of view of crystal structure, we find that the buckled-SiC$_2$-pentagon sheet has similar structure with the first three atomic layers of $\beta$$-$SiC(001)-c$(2\times2)$ SDB surface, as shown in \ref{fig6} (a). Referring to the successful preparation of graphene by mechanical exfoliation from graphite, we studied the possibility of cleaving the buckled-SiC$_2$-pentagon from the $\beta$-SiC(001)-$_c$$(2\times2)$ SDB surface through chemical exfoliation method with H/F implantation. Our calculations indicate that the first three atomic layers will spontaneously form the buckled SiC$_2$-pentagon when it is simply cleaved from $\beta$$-$SiC(001)-c$(2\times2)$ SDB surface into vacuum. To chemically exfoliate the first three atomic layer of the $\beta$$-$SiC(001)-c$(2\times2)$ SDB surface, we inject H atoms between the third and fourth layer of $\beta$$-$SiC(001)-c$(2\times2)$ SDB surface, as shown in \ref{fig6}(b). We find that the H atoms tend to bond with the electronegative C atom, such bonding effect induces the Si-C bonds broken between the third and fourth layer of $\beta$$-$SiC(001)-c$(2\times2)$ SDB surface. Then the first three layers of the $\beta$$-$SiC(001)-c$(2\times2)$ SDB surface is cleaved from the surface and the layer lifts up to 3{\AA} from the substrate even under equilibrium relaxation calculations. Such cleaved layer is with H saturation which could be reduced under annealing procedure. Besides, we also consider F atoms injecting into the $\beta$$-$SiC(001)-c$(2\times2)$ SDB surface, as shown in \ref{fig6} (c). The results show that when the injected F atom concentration is 1/2 layer, it doesn't form bond with the Si atom at hollow position of the $\beta$$-$SiC(001)-c$(2\times2)$ SDB surface. However, when doping concentration is 1 layer, F atoms bond to the Si atom at the hollow site of the $\beta$$-$SiC(001)-c$(2\times2)$  SDB surface, where the F-Si bond length is 1.6 {\AA} and the  $\angle$F-Si-F=78.693$^\circ$. The original Si-C bond between the third and fourth layer is borken, and the C atoms of the third layer form new bond with its neighboring Si atom, where all of the C-Si bond length is 2.124 {\AA} which is slightly longer than the bulk Si-C bond length (1.89 {\AA}). Moreover, the C atoms of the third layer not only form bond with the neighboring Si atom, but also form bond with each other with the bond length of 1.495{\AA}, as shown in \ref{fig6} (c). The configuration of the first three layer is like buckled SiC$_2$-pentagon structure adsorbed on the F decorated SiC(001) surface. The interaction between the first three layer SiC (is also called buckled SiC$_2$-pentagon) and the rest $\beta$$-$SiC(001)-c$(2\times2)$ SDB surface is greatly reduced by F atoms adsorption. We also calculate the electronic structure of  $\beta$$-$SiC(001)-c$(2\times2)$ SDB surface after the injection of H and F atoms. We can see that the band structures of  the first three SiC layers in F-injected configuration close to that of  buckled-SiC$_2$-pentagon, as shown in the \ref{fig5} (c). Our ab initio molecular dynamic (AIMD) simulation under the condition of 273.5 K within NVT shows that such buckled SiC$_2$-pentagon layer can be cleaved from the surface in the first 2 ps of our 10 ps simulation. In real experimental condition, such chemical exfoliation may produce buckled SiC$_2$-pentagon flakes rather than large scale sheet due to the complicated un-controllable procedure of the H/F implantation.  We look forward to the synthesis of buckled SiC$_2$-pentagon flakes or even sheet in experiments in the future to confirm our theoretical prediction.\\
\section{Conclusion}
\indent Using the segment combination method, we propose a novel two-dimensional nano-sheet of SiC$_2$ (buckled SiC$_2$-pentagon) consisting of tetrahedral silicon atoms
and triple-linked carbon atoms in a fully-pentagon network. With ab initio calculations, we find that the buckled SiC$_2$-pentagon is more favorable than its planar version (planar SiC$_2$-pentagon) and the previously proposed SiC$_2$ silagraphene with tetracoordinate silicon atoms. The buckled SiC$_2$-pentagon is an indirect-band-gap semiconductor with a gap of 1.388 eV and confirmed dynamically stable. Its one-dimensional nanoribbons can be metals or semiconductors depending on the edge types, width, and decorations. Interestingly, the buckled-SiC$_2$-pentagon might be synthesized through chemical exfoliation from $\beta$$-$SiC(001)-c$(2\times2)$ SDB surface. \\
\begin{acknowledgements}
This work is supported by the National Natural Science Foundation of China (Grant Nos. 10874143, 11274262, and 11274029 ), the Program for New Century Excellent Talents in University (Grant No. NCET-10-0169), the Natural Science Foundation of Hunan Province (Grant No. 13JJ4046), the Scientific Research Fund of Hunan Provincial Education Department (Grant No. 10K065) .
\end{acknowledgements}

\end{document}